# CHAPTER 1

# METAMATERIALS AND THE MATHEMATICAL SCIENCE OF INVISIBILITY

André Diatta, Sébastien Guenneau, André Nicolet, Fréderic Zolla

*Institut Fresnel (UMR CNRS 6133).*

*Aix-Marseille Université 13397 Marseille cedex 20, France*

E-mails: andre.diatta@fresnel.fr; sebastien.guenneau@fresnel.fr; andre.nicolet@fresnel.fr; frederic.zolla@fresnel.fr

Abstract.

In this chapter, we review some recent developments in the field of photonics: cloaking, whereby an object becomes invisible to an observer, and mirages, whereby an object looks like another one (say, of a different shape). Such optical illusions are made possible thanks to the advent of metamaterials, which are new kinds of composites designed using the concept of transformational optics. Theoretical concepts introduced here are illustrated by finite element computations.

## 1. Introduction

In the past six years, there has been a growing interest in electromagnetic metamaterials[1], which are composites structured on a subwavelength scale modeled using homogenization theories[2]. Metamaterials have important practical applications as they enable a markedly enhanced control of electromagnetic waves through coordinate transformations which bring anisotropic and heterogeneous[3] material parameters into their governing equations, except in the ray diffraction limit whereby material parameters remain isotropic[4]. Transformation[3] and conformal[4] optics, as they are now known, open an unprecedented avenue towards the design of such metamaterials, with the paradigms of invisibility cloaks.

In this review chapter, after a brief introduction to cloaking (section 2), we would like to present a comprehensive mathematical model of metamaterials introducing some basic knowledge of differential calculus. The touchstone of our presentation is that Maxwell's equations, the governing equations for electromagnetic waves, retain their form under coordinate changes. A modern formalism of differential calculus is that of differential forms (section 3), among which Withney forms offer a natural framework for finite elements in curvilinear coordinates. Illustrative numerical simulations are provided in section 4 (cloaking via a singular transform) and section 5 (cloaking via a non-singular transform). Section 6 summarizes the results presented in this chapter.

## 2. General aim of cloaking

In 2006, Pendry*et al.*[3] and Leonhardt[4] independently showed the possibility of designing a cloak that renders anyobject inside it invisible to electromagnetic radiation. This coating consists of a metamaterial whose physical properties, the electric permittivity and the magnetic permeability, are deduced from a coordinate transformation in the Maxwell system. The anisotropy and the heterogeneity of the parameters of the coat work as a deformation of the space around the object we want to hide by bending the wavefront around it and enabling waves to emerge on the other side in the original propagation direction without any perturbation, see Figure1 for the principle of cloaking. The experimental validation of these theoretical considerations was given, a few months later, by an international team involving the former authors who used a cylindrical cloak consisting of concentric arrays of split ring resonators. This cloak makes a copper cylinder invisible to an incident plane wave at 8.5 GHz[5] as predicted by the numerical simulations[6].

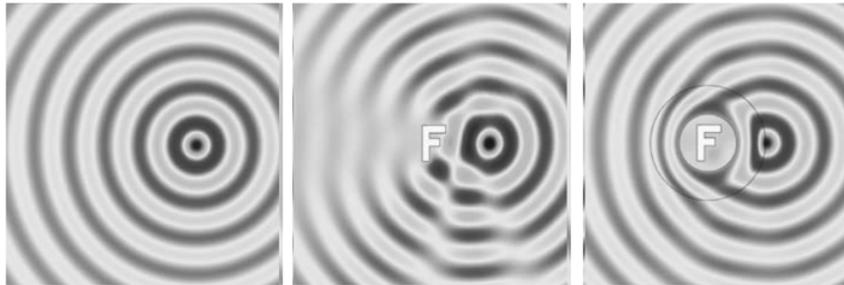

**Fig.1.** Principle of cloaking : (left) A point source radiating in an homogeneous isotropic medium; (middle) A point source radiating in an homogeneous isotropic medium in the presence of an infinite conducting F-shaped scatterer; (right) A point source radiating in an homogeneous isotropic medium in the presence of an infinite conducting F-shaped scatterer surrounded by an invisibility cloak (an anisotropic heterogeneous ring).

However, there are other types of cloaking, such as an invisibility cloak with a negative refractive index theorized by Milton and Nicorovici in 2006[7] and one with a low refractive index proposed by Alu and Engheta in 2005[8]. Such plasmonic routes to cloaking do not detour the wave trajectories, but rather counteract the scattering of an obstacle, e.g. the circular cylinder, with the resonance of a coating. We shall not discuss these alternatives to cloaking via geometric transforms, although they exist in their own right, and are equally interesting from the physical viewpoint.

## 3. Formalism of differential forms

One of the most appropriate frameworks for dealing with Maxwell Equations is the formalism of exterior calculus on differential forms. In such a setting, the magnetic field **H** and the electric field **E** are differential 1-forms, whereas the current density **j**, the magnetic flux density **B** and electric flux density **D** (also termed electric displacement) are differential 2-forms. The charge density $\rho$ is a 3-form. Maxwell Equations in the time-harmonic domain, encapsulate their relationship as follows

$$\mathbf{dH} = \mathbf{j} - i\omega\mathbf{D}, \quad \mathbf{dE} = i\omega\mathbf{B}, \quad \mathbf{dD} = \boldsymbol{\rho}, \quad \mathbf{dB} = \mathbf{0}. \qquad (3.1)$$

where **d** is the exterior derivative of differential forms[15].

Note that, in the first two and the last two relations of Eq. (3.1), the exterior derivative **d** plays the roles of **curl** and **div**, respectively. Note also that the fundamental property $\mathbf{dd} = \mathbf{0}$ always applies[15]. Remark that the relation $\mathbf{dE} = i\omega\mathbf{B}$ means that **B** coincides with the exact 2-form $\mathbf{B} = \mathbf{d}(-\frac{i}{\omega}\mathbf{E})$ and hence implies, in particular, the relation $\mathbf{dB} = 0$. Furthermore, the equation $\mathbf{dH} = \mathbf{j} - i\omega\mathbf{D}$ also implies that $\mathbf{dj} = i\omega\mathbf{dD} + \mathbf{d}^2\mathbf{H} = i\omega\mathbf{dD} = i\omega\boldsymbol{\rho}$.

Now consider a metric **g** on, say $\mathbb{R}^n$, that is, a nondegenerate symmetric bilinear map from the space of vector fields on $\mathbb{R}^n$ to the space of smooth scalar valued functions. More precisely, two vectors $\mathbf{V}_p$ and $\mathbf{V}_p^{'}$ originating from a point p are mapped into a scalar $\mathbf{g}(\mathbf{V}_p, \mathbf{V}_p^{'})$. The bilinearity means that this correspondence depends linearly on both arguments $\mathbf{V}_p$ and $\mathbf{V}_p^{'}$. Making the point p vary over $\mathbb{R}^n$, or equivalently, using vector fields $\mathbf{V}, \mathbf{V}^{'}$, gives rise to a smooth function $\mathbf{g}(\mathbf{V}, \mathbf{V}^{'})$ on $\mathbb{R}^n$.

The Hodge star operator $*$ associated to the metric **g**, is a linear operator on differential forms, bijectively mapping the space of p-forms to that of (n-p)-forms on $\mathbb{R}^n$, for every fixed p which could be 0, 1, 2,… or n. It is a very convenient tool, allowing to define a scalar product

$$(\boldsymbol{\alpha}, \boldsymbol{\beta}) := \int_{\mathbb{R}^n} \boldsymbol{\alpha} \wedge *\boldsymbol{\beta} \qquad (3.2)$$

on the space of differential p-forms, on $\mathbb{R}^n$. In the above equations (3.2), the symbol $\wedge$ stands for the (skew-symmetric) exterior (wedge) product on differential forms. It could be shown that, endowed with the scalar product (3.2), the space of differential p-forms on $\mathbb{R}^n$ becomes a Hilbert space. This property is then used here, in a weak formulation of Maxwell's Equations, to derive the magnetic permeability and the electric permittivity coefficients of a transformed medium. To keep things simple, we will also restrict ourselves to the case where **g** is the Euclidean metric. Written in Cartesian coordinates $(x_1,...,x_n)$ with corresponding local basis $(\frac{\partial}{\partial x_1},...,\frac{\partial}{\partial x_n})$ for vector fields, the metric has a symmetric matrix with (i,j) entry $g_{ij} = \mathbf{g}(\frac{\partial}{\partial x_i},\frac{\partial}{\partial x_j}) = \delta_{ij}$, where $\delta_{ij}$ are the Kronecker symbols. In our case of interest here, n=3 and p=0, 1, 2 or 3. We endow $\mathbb{R}^3$ with the Euclidean metric and denote its usual global coordinates by $(x_1, x_2, x_3) = (x, y, z)$, in this case we have

$$*\mathbf{1} = \mathbf{dx} \wedge \mathbf{dy} \wedge \mathbf{dz},$$
$$*\mathbf{dx} \wedge \mathbf{dy} \wedge \mathbf{dz} = \mathbf{1},$$
$$*\mathbf{dx} = \mathbf{dy} \wedge \mathbf{dz},$$
$$*\mathbf{dy} \wedge \mathbf{dz} = \mathbf{dx},$$
$$*\mathbf{dy} = \mathbf{dz} \wedge \mathbf{dx}, \quad (3.3)$$
$$*\mathbf{dz} \wedge \mathbf{dx} = \mathbf{dy},$$
$$*\mathbf{dz} = \mathbf{dx} \wedge \mathbf{dy},$$
$$*\mathbf{dx} \wedge \mathbf{dy} = \mathbf{dz}.$$

Let us recall that the differential exact 1-forms **dx**, **dy**, **dz** form the (local) basis for covectors, dual to $(\frac{\partial}{\partial x},\frac{\partial}{\partial y},\frac{\partial}{\partial z})$. Using (3.3), we can derive the following interesting property of the Hodge star operator. If $\boldsymbol{\alpha} = \alpha_x\mathbf{dx} + \alpha_y\mathbf{dy} + \alpha_z\mathbf{dz}$ and $\boldsymbol{\beta} = \beta_x\mathbf{dx} + \beta_y\mathbf{dy} + \beta_z\mathbf{dz}$ are two differential 1-forms on $\mathbb{R}^3$, then the following holds

$$\boldsymbol{\alpha} \wedge *\boldsymbol{\beta} = (\alpha_x\beta_x + \alpha_y\beta_y + \alpha_z\beta_z)\ \mathbf{dx} \wedge \mathbf{dy} \wedge \mathbf{dz}$$
$$= (\alpha_x, \alpha_y, \alpha_z)(\beta_x, \beta_y, \beta_z)^T\ \mathbf{dx} \wedge \mathbf{dy} \wedge \mathbf{dz}. \quad (3.4)$$

Next, we use the following, which we may call the change of volume rule under change of coordinates. Suppose we have another coordinate system $(u, v, w)$, with the following change of coordinates $(x, y, z) = (x(u, v, w), y(u, v, w), z(u, v, w))$. The Jacobian matrix of this change of coordinates is

$$\mathbf{J}(u,v,w) = \begin{pmatrix} \dfrac{\partial x}{\partial u} & \dfrac{\partial x}{\partial v} & \dfrac{\partial x}{\partial w} \\ \dfrac{\partial y}{\partial u} & \dfrac{\partial y}{\partial v} & \dfrac{\partial y}{\partial w} \\ \dfrac{\partial z}{\partial u} & \dfrac{\partial z}{\partial v} & \dfrac{\partial z}{\partial w} \end{pmatrix}$$

Using the determinant of the above matrix, the formula below indicates how the volume changes as we switch from a system of coordinates to another:

$$\mathbf{dx} \wedge \mathbf{dy} \wedge \mathbf{dz} = \det(\mathbf{J})\, \mathbf{du} \wedge \mathbf{dv} \wedge \mathbf{dw}. \tag{3.5}$$

The relations

$$\mathbf{dx} = \frac{\partial x}{\partial u}\mathbf{du} + \frac{\partial x}{\partial v}\mathbf{dv} + \frac{\partial x}{\partial w}\mathbf{dw}, \quad \mathbf{dy} = \frac{\partial y}{\partial u}\mathbf{du} + \frac{\partial y}{\partial v}\mathbf{dv} + \frac{\partial y}{\partial w}\mathbf{dw},$$

$$\mathbf{dz} = \frac{\partial z}{\partial u}\mathbf{du} + \frac{\partial z}{\partial v}\mathbf{dv} + \frac{\partial z}{\partial w}\mathbf{dw},$$

could be written in a compact way as

$$\begin{pmatrix} \mathbf{dx} \\ \mathbf{dy} \\ \mathbf{dz} \end{pmatrix} = \mathbf{J} \begin{pmatrix} \mathbf{du} \\ \mathbf{dv} \\ \mathbf{dw} \end{pmatrix}.$$

Now we can also deduce the relation between the expressions of a given 1-form $\boldsymbol{\alpha}$ in the two coordinate systems, as follows

$$\boldsymbol{\alpha} = \alpha_x \mathbf{dx} + \alpha_y \mathbf{dy} + \alpha_z \mathbf{dz}$$

$$= (\alpha_x, \alpha_y, \alpha_z) \begin{pmatrix} \mathbf{dx} \\ \mathbf{dy} \\ \mathbf{dz} \end{pmatrix} = (\alpha_x, \alpha_y, \alpha_z) \mathbf{J} \begin{pmatrix} \mathbf{du} \\ \mathbf{dv} \\ \mathbf{dw} \end{pmatrix} \quad (3.6)$$

On the other hand, we could also write directly

$$\boldsymbol{\alpha} = \alpha_u \mathbf{du} + \alpha_v \mathbf{dv} + \alpha_w \mathbf{dw} = (\alpha_u, \alpha_v, \alpha_w) \begin{pmatrix} \mathbf{du} \\ \mathbf{dv} \\ \mathbf{dw} \end{pmatrix}. \quad (3.7)$$

Thus we get the equality

$$(\alpha_u, \alpha_v, \alpha_w) = (\alpha_x, \alpha_y, \alpha_z) \mathbf{J} \quad (3.8)$$

From Eqs. (3.5) and (3.7), we can rewrite (3.4) and see how the 3-form $\boldsymbol{\alpha} \wedge *\boldsymbol{\beta}$ is transformed under the change of coordinates:

$$\boldsymbol{\alpha} \wedge *\boldsymbol{\beta} = (\alpha_u, \alpha_v, \alpha_w) \mathbf{J}^{-1} ((\beta_u, \beta_v, \beta_w) \mathbf{J}^{-1})^T \det(\mathbf{J}) \, \mathbf{du} \wedge \mathbf{dv} \wedge \mathbf{dw}$$
(1.3)
That is,
$$\boldsymbol{\alpha} \wedge *\boldsymbol{\beta} = (\alpha_u, \alpha_v, \alpha_w) \mathbf{J}^{-1} \mathbf{J}^{-T} \det(\mathbf{J}) \, (\beta_u, \beta_v, \beta_w)^T \, \mathbf{du} \wedge \mathbf{dv} \wedge \mathbf{dw} \quad (3.9)$$

This is easily generalized to more general piecewise differentiable maps between two domains.

Now, for simplicity, we place ourselves in a region where there are no charges or currents. Under these considerations, Eq. (3.1) now reads

$$\mathbf{dH} = i\omega \mathbf{D}, \quad \mathbf{dE} = i\omega \mathbf{B}, \quad \mathbf{dD} = \mathbf{0}, \quad \mathbf{dB} = \mathbf{0}.$$

For example, we can eliminate $\mathbf{H}$ in the equation $\mathbf{dH} = i\omega \mathbf{D}$ and get the wave equation of the electric field $\mathbf{E}$ in homogeneous media, and using the Hodge operator, we get

$$\mathbf{d}(\mu^{-1} * \mathbf{dE}) - \omega^2 \varepsilon * \mathbf{D} = \mathbf{0}.$$

Here, the electric field **E** is seen as an unknown. The above allows us to use the scalar product defined in Eq. (3.2) to write the weak formulation as follows

$$\int_\Omega \mu^{-1} * \mathbf{dE} \wedge \mathbf{d\dot{E}'dx} - \omega^2 \int_\Omega \varepsilon * \mathbf{E} \wedge \mathbf{\dot{E}'dx} = \mathbf{0}, \qquad (3.10)$$

for every element **E'** of $H_0(\mathbf{curl}, \Omega)$.

In an isotropic medium, the electric permittivity $\varepsilon$ and the electromagnetic permeability $\mu$ are scalar valued functions (indeed of the form of a scalar function times the identity matrix). So starting our transformation from an isotropic medium occupying a domain D, the above equalities allow us to extract the following relations between the expressions of the electric permittivity and the magnetic permeability in the different coordinate systems[18].

$$\boldsymbol{\varepsilon'} = \varepsilon \mathbf{T}^{-1} \text{ and } \boldsymbol{\mu'} = \mu \mathbf{T}^{-1}, \text{ where } \mathbf{T} = \frac{\mathbf{J}^T \mathbf{J}}{\det(\mathbf{J})}. \qquad (3.11)$$

Remark that, we have started from an isotropic and homogeneous medium, so that ε and μ are constant numbers. And yet, because of the structure of the tensor **T** in the transformed medium, we end up having tensors **ε'** and **μ'** depending on the space variable. Hence, the transformed medium is always highly anisotropic and inhomogeneous.

It could be checked, using exactly the same process, that if we start from anisotropic and inhomogeneous medium with tensors **ε** and **μ**, then (3.10) takes the form

$$\boldsymbol{\varepsilon'} = \mathbf{J}^{-1}\boldsymbol{\varepsilon}\mathbf{J}^{-T}\det(\mathbf{J}) \text{ and } \boldsymbol{\mu'} = \mathbf{J}^{-1}\boldsymbol{\mu}\mathbf{J}^{-T}\det(\mathbf{J}), \qquad (3.12)$$

which defines an anisotropic and inhomogeneous transformed medium.

Note that, one can write the transformation in any coordinate systems of their choice, and to get back to Cartesian ones in order to use formulas (3.11) or (3.12), one just needs to perform a composition of the elementary Jacobians involved, as in Eq. (4.4) of Section 4.2.

## 4. Cloaking with a singular transformation

In this section, we discuss the physics of cloaking via a singular transform blowing up a point onto a disc of radius $R_1$, while mapping a disc of radius $R_2$ onto itself. We note that this transform was introduced in 2003 by Greenleaf et al.[9] in the context of inverse problems for the conductivity equation.

### 4.1. Invisibility

Using the same notations as in Section 3, we show in Figure 2 the effect of the previous transform on the metric of space. The distorted metric is associated with an anisotropic heterogeneous medium defined through the Jacobian **J** of the transform via the formula (3.11). Such an invisibility cloak is defined by tensors of permittivity and permeability with extreme coefficients at the inner radius of the cloak (vanishing and infinite eigenvalues). It leads to perfect cloaking, as numerically illustrated in Figure 1 (right). However, one can also give a twist to such a cloak and use it to create a mirage effect[6,10], as we discuss in the next paragraph.

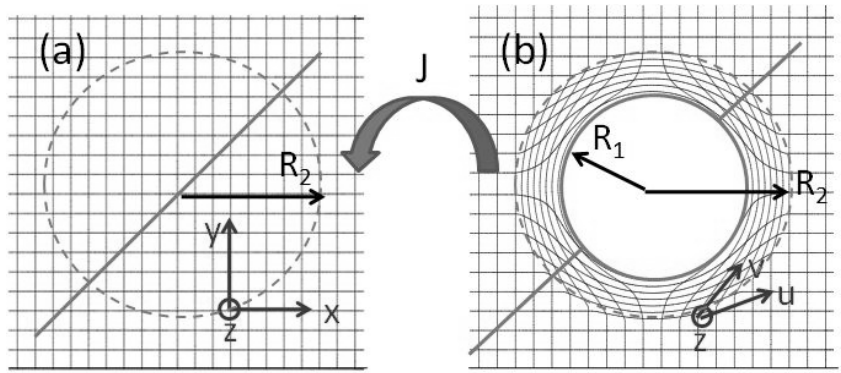

**Fig.2.** Effect of singular transform on space metric: (a) Metric of space with coordinate system (x,y,z) for a homogeneous isotropic medium; (b) Metric of space with coordinate system (u,v,z) for a locally heterogeneous anisotropic medium inside the invisibility cloak which is an annulus of inner radius $R_1$ and outer radius $R_2$.

### 4.2 Mirage effect

Transformation Optics/Electromagnetics heavily relies on the equivalence principle for Electromagnetics stating that, any electromagnetic medium creates geometry and vice versa. The electromagnetic properties of such a medium entirely characterize the metric (including non-Riemannian metrics, such as spacetime metrics, depending on the cases) of the geometry. The converse also holds true. The trick in generalizing Pendry's cloak and using it as a means to produce optical illusions and mimesis also falls along these lines. The aim of this section is expose a proposal initiated by two of us[10] to show that not only the whole (circular) region bounded by $\|\mathbf{x}'\|= R_1$ is equivalent to an optical void, but any defect hidden inside the cloak $R_1 < \|\mathbf{x}'\| < R_2$ will automatically acquire the same optical/electromagnetic properties as its inverse image via the transformation used to design the cloak, see Figure 1. Any incoming wave will see such an object as its counterpart (inverse) placed in vacuum, and will bear this resemblance to an observer. The transformation used maps $\mathbf{x} = (x, y, z)$ to $\mathbf{x}' = (x', y', z')$ is given by

$$\mathbf{x}' = \left( \frac{R_2 - R_1}{R_2} + \frac{R_1}{\|\mathbf{x}\|} \right) \mathbf{x}, \qquad (4.1)$$

where $\|\mathbf{x}\|$ is the norm (modulus) of $\mathbf{x}$.

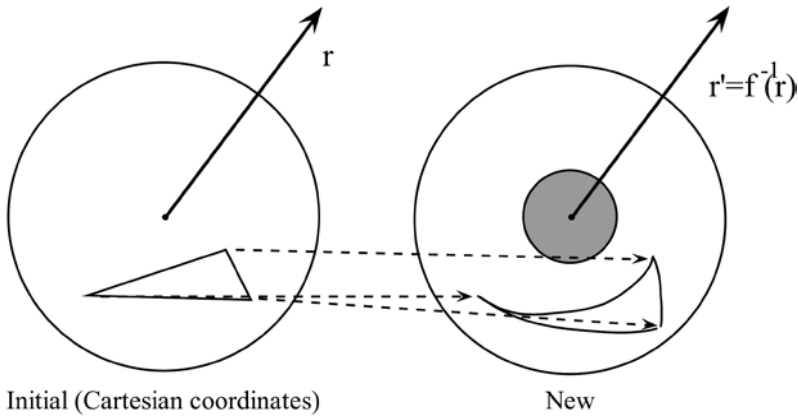

Initial (Cartesian coordinates)          New

**Fig.3.** Construction of the transformation $f : \mathbf{r} = (x, y, z) \mapsto \mathbf{r}' = (x', y', z')$ giving rise to the generalized cloak. When the material properties are piecewise defined, a pushforward of the geometry involving the inverse transformation is useful.

Now we can choose to use cylindrical coordinates $(r, \theta, z)$ which are related Cartesian coordinates $(x, y, z)$ by

$$x = r\cos(\theta), y = r\sin(\theta), \quad z = z$$

with corresponding Jacobian matrix

$$\mathbf{J}_{xr}(r,\theta,z) = \begin{pmatrix} \cos(\theta) & -r\sin(\theta) & 0 \\ \sin(\theta) & r\cos(\theta) & 0 \\ 0 & 0 & 1 \end{pmatrix} = \begin{pmatrix} \cos(\theta) & -\sin(\theta) & 0 \\ \sin(\theta) & \cos(\theta) & 0 \\ 0 & 0 & 1 \end{pmatrix} \begin{pmatrix} 1 & 0 & 0 \\ 0 & r & 0 \\ 0 & 0 & 1 \end{pmatrix}.$$

Note the appearance of the matrix of rotation through an angle $\theta$ whose inverse is the rotation through an angle $-\theta$. The inverse of this Jacobian is then easy to compute

$$\mathbf{J}_{xr}^{-1} = \mathbf{J}_{rx} = \begin{pmatrix} 1 & 0 & 0 \\ 0 & \frac{1}{r} & 0 \\ 0 & 0 & 1 \end{pmatrix} \begin{pmatrix} \cos(\theta) & \sin(\theta) & 0 \\ -\sin(\theta) & \cos(\theta) & 0 \\ 0 & 0 & 1 \end{pmatrix} = \begin{pmatrix} \cos(\theta) & \sin(\theta) & 0 \\ -\frac{1}{r}\sin(\theta) & \frac{1}{r}\cos(\theta) & 0 \\ 0 & 0 & 1 \end{pmatrix}$$

We may then rewrite the transformation (4.1) as mapping $(r, \theta, z)$ to new coordinates $(r', \theta', z')$. It maps $\mathbb{R}^3$ to $\mathbb{R}^3$ minus a cylinder over a disk $D_1$ of radius $R_1$, in such a way as it is the identity outside a cylinder over a disk $D_2$ of radius $R_2 > R_1$ and inside the cylindrical hollow region over the annulus $R_1 < r \leq R_2$, it reads

$$r' = \frac{R_2 - R_1}{R_2}r + R_1, \quad \theta' = \theta, \quad z' = z, \tag{4.2}$$

so that the origin is blown up to the circle of radius $R_1$. All discs and circles referred to here, are contained in the xy-plane and centered at the origin. Setting $\alpha = \dfrac{R_2 - R_1}{R_2}$, the Jacobian of this transformation is

$$\mathbf{J}_{r'r}(r,\theta,z) = \begin{pmatrix} \alpha & 0 & 0 \\ 0 & 1 & 0 \\ 0 & 1 & 1 \end{pmatrix} \qquad (4.3)$$

To deduce the Jacobian in Cartesian coordinates, we compose the three elementary Jacobians as follows

$$\mathbf{J} := \mathbf{J}_{x'x} = \mathbf{J}_{x'r'}\mathbf{J}_{r'r}\mathbf{J}_{rx} \qquad (4.4)$$

so that, we have

$$\mathbf{J}^{-1} = (\mathbf{J}_{rx})^{-1}(\mathbf{J}_{r'r})^{-1}(\mathbf{J}_{x'r'})^{-1} = \mathbf{J}_{xr}\mathbf{J}_{rr'}\mathbf{J}_{r'x'}$$

$$= \begin{pmatrix} \cos(\theta) & -r\sin(\theta) & 0 \\ \sin(\theta) & r\cos(\theta) & 0 \\ 0 & 0 & 1 \end{pmatrix} \begin{pmatrix} \dfrac{1}{\alpha} & 0 & 0 \\ 0 & 1 & 0 \\ 0 & 0 & 1 \end{pmatrix} \begin{pmatrix} \cos(\theta) & \sin(\theta) & 0 \\ -\dfrac{1}{r'}\sin(\theta) & \dfrac{1}{r'}\cos(\theta) & 0 \\ 0 & 0 & 1 \end{pmatrix}$$

We are now ready to use formula (3.11) in order to derive the electromagnetic properties of the cloak. We get

$$\mathbf{T}^{-1} = \begin{pmatrix} (\mathbf{T}^{-1})_{11} & (\mathbf{T}^{-1})_{12} & 0 \\ (\mathbf{T}^{-1})_{21} & (\mathbf{T}^{-1})_{22} & 0 \\ 0 & 0 & \dfrac{r' - R_1}{\alpha^2 r'} \end{pmatrix} \qquad (4.5)$$

where the coefficients are given as follows

$$(\mathbf{T}^{-1})_{11} = 1 - \frac{R_1 \sin^2(\theta')}{r'} + \frac{R_1 \cos^2(\theta')}{r' - R_1},$$

$$(\mathbf{T}^{-1})_{22} = 1 - \frac{R_1 \cos^2(\theta')}{r'} + \frac{R_1 \sin^2(\theta')}{r' - R_1}, \qquad (4.6)$$

$$(\mathbf{T}^{-1})_{12} = (\mathbf{T}^{-1})_{21} = \frac{R_1 \cos(\theta')\sin(\theta')(R_1 - 2r')}{(R_1 - r')r'}.$$

We can write these expressions only in terms of $(x', y', z')$, using

$$r' = \sqrt{(x')^2 + (y')^2}, \qquad \theta' = 2 \arctan\left(\frac{y'}{x' + \sqrt{(x')^2 + (y')^2}}\right), \qquad z' = z.$$

The electromagnetic permeability **μ** and electric permittivity **ε** of the cloak are then ready to be implemented, applying (3.11).

Now following Nicolet et al.[10] we can apply the transformation (4.2), but this time, the disc $D_2$ contains an object of arbitrary electromagnetic properties. The cloaks now mimics the object originally contained in $D_2$. A numerical simulation with the Getdp freeware is shown in Figure 4.

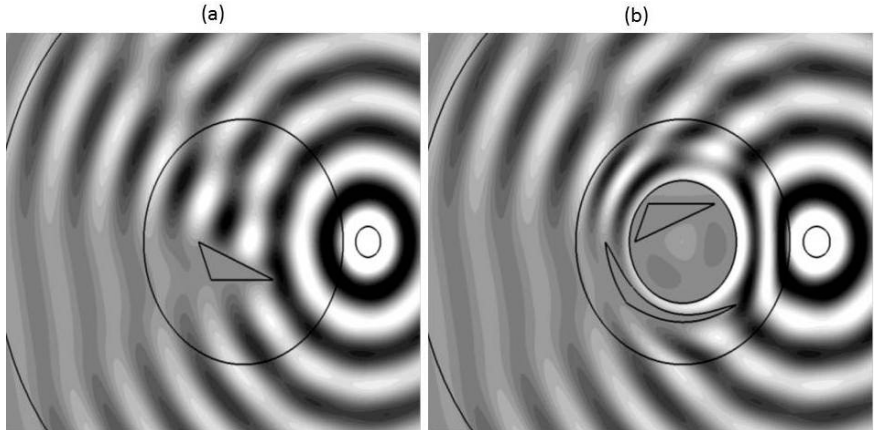

**Fig.4.**

(a) Conducting triangular cylinder scattering cylindrical waves radiated by an electric line source (white disc with a black perimeter on the right-hand side) located nearby.

(b) Distorted conducting cylinder placed inside a cloak $R_1 < ||\mathbf{x}'|| < R_2$, producing the same scattering pattern as in panel (a). Note that the triangular cylinder placed inside the invisibility region $||\mathbf{x}'|| \leq R_1$ does not affect the scattering, and could have in principle any desired shape and constituting material.

## 5. Cloaking and Mirage effect with a non-singular transformation

The mathematical model given below (and proposed by two of us[12]) provides a general setting for the design of a cloak aimed at concealing a region having the property that all of its points are within line-of-sight from a fixed vantage point. We call a star domain, every domain with such a property. Discs and Squares in 2D, ellipsoids in 3D … are examples of star domains. The designed cloaks play the double role of displaying any desired optical effect, in particular acquiring the same optical properties as other different objects on the one hand, as well as allowing to hide any defects inside it, just as an ordinary cloak would do, on the other hand. The mathematical model underlying this design is given by a nonsingular transformation mapping the two domains (objects to be made equivalent) on each other. Such a transformation is made to preserves every line passing through the chosen vantage point. It is a more general piecewise smooth and nonsingular transformation sending a shape to another one of a different geometry. Depending on whether objects to be mimicked live in isotropic or anisotropic media, formulae (3.11) and (3.12) will be used, accordingly. However, as noted in Section 3, the obtained cloak itself will always be a heterogeneous and anisotropic medium and yet will look to an external observer just as the object we wish to mimic. For the sake of simplicity in applications, we will consider here that, our imaginary objects to be mimicked are surrounded by an isotropic region, so that the permeability $\mu'$ and permittivity $\varepsilon'$ can be implemented using (3.11), to make the two objects acquire equivalent electromagnetic properties.

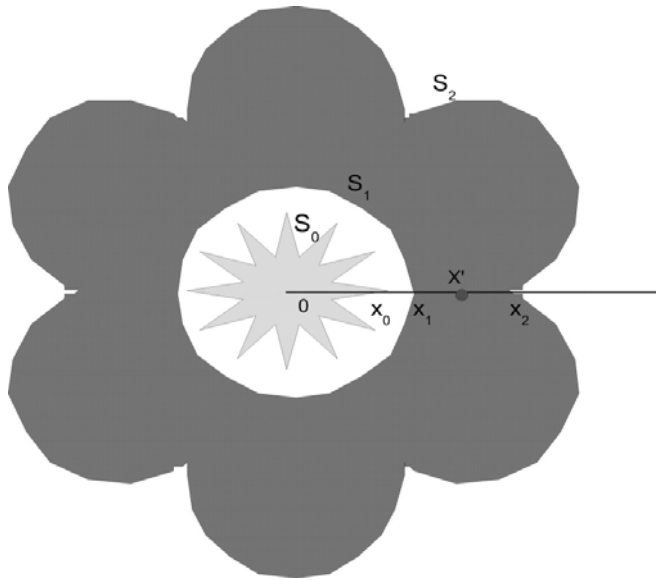

**Fig. 5.** Construction of a generalized non-singular cloak for mirage effect. The transformation with inverse (5.1) shrinks the region bounded by the two surfaces $S_0$ and $S_2$ into the region bounded by $S_1$ and $S_2$. The curvilinear metric inside the carpet (here, a dark gray flower) is described by the transformation matrix T, as in (3.11). This is designed to play the double role of mimesis and cloaking: any types of objects located within the region $D_1$ bounded by the surface S1 will be invisible to an outer observer while the region itself still scatters waves like an object $D_0$ bounded by $S_0$ (here, a star). In the limit of vanishing light gray region, the transformation matrix T becomes singular on S1 (ordinary invisibility cloak).

Here is a short mathematical description of the transformation and the derivation of the material properties of the designed cloak. Suppose three bounded star domains $D_0$, $D_1$, $D_2$ in $\mathbb{R}^n$ with, say, n=2,3,..., have piecewise smooth arbitrary boundaries $\partial D_j$, j=0,1,2. We chose them nested, that is, $D_2$ contains $D_1$ which in turn, contains $D_0$ and they share the same vantage point, say $\underline{\mathbf{0}}$.

The domain $D_0$ is our reference, the one to be mimicked by $D_1$ and thereby by the cloak itself. The hollow region $D_2 \setminus D_1$ will be endowed with Neumann conditions on its inner boundary $\partial D_1$. It is meant to be the model for the cloak and in which any type of defect could be hidden, but will still have the same electromagnetic response as the region $D_0$ with an infinitely conducting boundary $\partial D_0$.

The transformation coincides with the identity map outside $D_2$, that is, in $\mathbb{R}^n \setminus D_2$, but bijectively (indeed, at least, piece-wise diffeomorphically) maps the hollow region $D_2 \setminus D_0$ to the hollow region $D_2 \setminus D_1$, so that $\partial D_2$ stays point-wise fixed, while $\partial D_0$ is mapped to $\partial D_1$. We may divide the domains $D_j$ into subdomains, such that each part of the boundaries lying inside $\partial D_j$ is an arbitrary smooth hypersurface.

Now for the construction of the transformation, let $\underline{\mathbf{x}} = (x^1, x^2, ..., x^n)$ be a point in $D_1 \setminus D_0$, whose coordinates are given in a system of coordinates centered at the chosen vantage point $\underline{\mathbf{0}}$. A line segment issued from $\underline{\mathbf{0}}$ and passing through $\underline{\mathbf{x}}$, meets the boundaries $\partial D_0, \partial D_1, \partial D_2$ at the unique points
$\underline{\mathbf{x}}_0 = (x_0^1, x_0^2, ..., x_0^n)$, $\underline{\mathbf{x}}_1 = (x_1^1, x_1^2, ..., x_1^n)$, $\underline{\mathbf{x}}_2 = (x_2^1, x_2^2, ..., x_2^n)$, respectively. For numerical implementations, we directly use the inverse $\underline{\mathbf{x}}' \mapsto \underline{\mathbf{x}}$ of the transformation, in the coordinates system. It is given by

$$x^i = x_0^i + \alpha_i \ (x'^i - x_1^i), \text{ where } \alpha_i = \frac{x_2^i - x_0^i}{x_2^i - x_1^i}. \tag{5.1}$$

In the 3-space with coordinates $(x^1, x^2, x^3) = (x, y, z)$ we can write this transformation as

$$\begin{cases} x = x_0 + \alpha \ (x' - x_1), \text{ with } \alpha = \dfrac{x_2 - x_0}{x_2 - x_1} \\ y = y_0 + \beta \ (y' - y_1) \text{ with } \beta = \dfrac{y_2 - y_0}{y_2 - y_1} \\ z = z_0 + \gamma \ (z' - z_1), \text{ with } \gamma = \dfrac{z_2 - z_0}{z_2 - z_1} \end{cases} \tag{5.2}$$

The cylindrical cases (cylinders over plane curves, such as triangles, squares, ellipses, sun flower-like cylinders, and so forth) are given by the following transformation mapping the region enclosed between the cylinders $S_0$ and $S_2$ into the space between $S_1$ and $S_2$. Its inverse reads

$$\begin{cases} x = x_0 + \alpha \ (x' - x_1), \text{ with } \alpha = \dfrac{x_2 - x_0}{x_2 - x_1} \\ y = y_0 + \beta \ (y' - y_1) \text{ with } \beta = \dfrac{y_2 - y_0}{y_2 - y_1} \\ \phantom{xxxxx} z = z'. \end{cases}$$

(5.3)

Applying formula (3.11), we derive the material properties, via the matrix representation of the tensor $\mathbf{T}^{-1}$, which now has the following simpler form[12]

$$\mathbf{T}^{-1} = \begin{pmatrix} T_{11}^{-1} & T_{12}^{-1} & 0 \\ T_{12}^{-1} & T_{22}^{-1} & 0 \\ 0 & 0 & T_{33}^{-1} \end{pmatrix} \tag{5.4}$$

with coefficients

$$T_{11}^{-1} = \frac{a_{11}}{a_{33}}, \; T_{12}^{-1} = \frac{a_{12}}{a_{33}}, \; T_{22}^{-1} = \frac{a_{22}}{a_{33}}, \; T_{33}^{-1} = a_{33} \qquad (5.5)$$

where the functions $a_{ij}$ are given by

$$a_{11} = \left(\frac{\partial x}{\partial y'}\right)^2 + \left(\frac{\partial y}{\partial y'}\right)^2, \; a_{12} = -\left(\frac{\partial x}{\partial x'}\frac{\partial x}{\partial y'} + \frac{\partial y}{\partial x'}\frac{\partial y}{\partial y'}\right),$$

$$a_{22} = \left(\frac{\partial x}{\partial x'}\right)^2 + \left(\frac{\partial y}{\partial x'}\right)^2, \; a_{33} = \frac{\partial x}{\partial x'}\frac{\partial y}{\partial y'} - \frac{\partial x}{\partial y'}\frac{\partial y}{\partial x'}. \qquad (5.6)$$

All we have to do now, is to plug in the specific values of all those $a_{ij}$ and then obtain $\boldsymbol{\mu}'$ and $\boldsymbol{\varepsilon}'$, via formulae (3.11). The form of the functions $x_i$, i=0,1,2 depend on the shapes of the regions $D_i$ involved in the cloaking process. So are the $a_{ij}$.

### 5.1. Squaring the circle

Following the above, we may want to make a circle acquire the same (electromagnetic) signature as a virtual small square sitting inside its enclosed region and sharing the same centre used as a vantage point[12]. By doing so, such a circle will have the same appearance to an observer just as a small square. Consider the two diagonals of the square, they part the region enclosed between the small square and the outer circle into four sectors. In each sector, following segments issued from the origin (common centre of both the square and circle), the transformation shrinks diffeomorphically the region between the small square and the outer circle into the circular annulus bounded by the inner and outer circles, in such a way as the sides of the square are mapped to the inner circle and the outer circle stays fixed point-wise.
First, let us recall[12] that in 2D, if a piece of the boundary of a star domain $D_i$ is part of a line of the form $y = a_i x + b_i$, then clearly the line through the origin and a point $(x', y')$ intersects this piece of boundary at

$$(x_i, y_i) = (\frac{b_i x'}{y' - a_i x'}, \frac{b_i y'}{y' - a_i x'}) \quad (5.7)$$

or, if this piece of boundary is a vertical segment with equation $x = c_i$, at

$$(c_i, c_i \frac{y'}{x'}). \quad (5.8)$$

In the case where this piece of boundary is part of an ellipse with equation

$$\frac{(x-a)^2}{c_i^2} + \frac{(y-b)^2}{d_i^2} = 1$$

then[12], a segment issued from $(a,b)$ passing through $(x', y')$ will intersect such a piece of boundary at $(x_i, y_i)$ with[12]

$$x_i = a + \frac{x' - a}{\sqrt{(x'-a)^2 / c_i^2 + (y'-a)^2 / d_i^2}},$$

$$y_i = a + \frac{y' - a}{\sqrt{(x'-a)^2 / c_i^2 + (y'-a)^2 / d_i^2}} \quad (5.9)$$

For the construction of our cloak (see Figure 6,) we use for $D_0$ a region bounded by a square with sides of length $L_0 = 0.2$ centred at the vantage point $(a,b) = (0,0)$ so that in both the leftmost and rightmost parts of Figure 6, formula (5.8) applies to give

$$(x_0, y_0) = (L_0, L_0 \frac{y'}{x'}) \quad (5.10)$$

whereas in both the uppermost and lowermost parts, we will invoke formula (3.7) to obtain

$$(x_0, y_0) = (L_0 \frac{x'}{y'}, L_0). \quad (5.11)$$

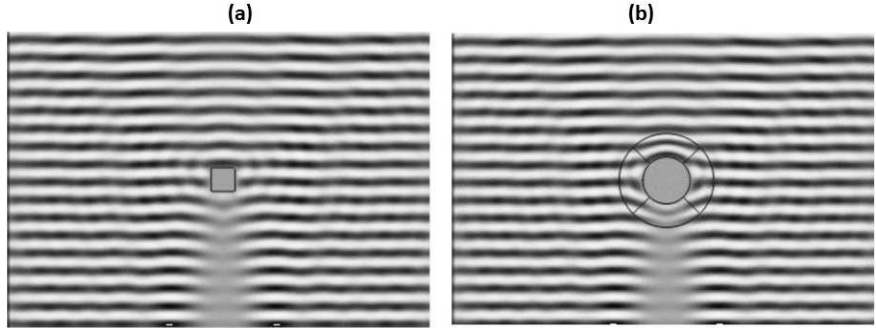

**Fig. 6.** A plane wave incident from above on (a) a small perfectly conducting square cylinder and (b) a circular nonsingular cloak is scattered in exactly the same way.

For both regions $D_1$ and $D_2$, we use two discs, bounded by circles with radii $R_1 = c_1 = d_1 = 0.2$, $R_2 = c_2 = d_2 = 0.4$, respectively and all centered at $(a,b) = (0,0)$. So here formulae (5.9) apply in all sectors to give

$$(x_i, y_i) = (R_i \frac{x'}{\sqrt{x^2 + y^2}}, R_i \frac{y'}{\sqrt{x^2 + y^2}}), \text{ i=1,2}. \quad (5.12)$$

Plugging (5.10)-(5.12) in (5.3) then in (5.5)-(5.6), we explicitly obtain $\mathbf{T}^{-1}$ using (5.9) and readily deduce the material properties of the cloak (annulus of radii $R_1$ and $R_2$) via formulae (3.11) as shown in Figure 6.

### 5.2. Optical illusions with squares

In this section, we apply exactly the same procedure as in Section 5.1 in order to design square-like cylindrical cloaks that acquire the same optical signature as an as smaller square-like cylinder as we want. Hence such cloaks will in particular appear to external observer as a very small cylinder. Here, all the domains involved have linear boundaries, so that we may chose formulas (5.7) and (5.8) as shown in Figure 7. The same algorithm can be used to design any nonsingular generalized cloak of this kind, where the domains involved have a mixture of elliptic and/or linear boundaries. Such a cloak will then make cross-like, sunflower-like, polygram-like, ellipse-like, … objects have the same electromagnetic signature as one another and in particular look like one another to an external observer.

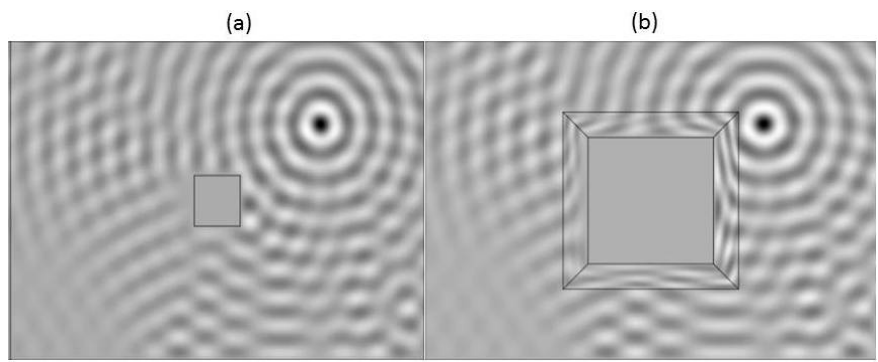

**Fig. 7. (a)** Scattering by a point source in the presence of a small infinite conducting square obstacle. **(b)** Scattering by a point source in the presence of a large infinite conducting square obstacle. The scattered fields outside the region occupied by the large square obstacle are identical in (a) and (b).

## 6. Conclusion

In this chapter, some recent developments in the theory of cloaking, using metamaterials, are presented. In particular, in sections 4.2 and 5. we discussed two different approaches to mirage effects in invisibility cloaking.

The design of mimesis cloaks we discussed in Section 5 (see Diatta and Guenneau[12]), only involves nonsingular transformations which are determined by the optical properties we would like to confer to our cloaked objects. It maps the two objects to be made equivalent, in a one-to-one smooth way. In this way, any object can acquire the same optical properties as any chosen other one. However, such a design requires Neumann conditions on the inner boundaries.

In contrast, the generalized cloaking technique as in Section 4, generally involves a singular transformation that gives a given object the appearance of any desired one without requiring Perfect Conducting boundary (Neumann) conditions. It is an extension of the mirage effect discussed in Nicolet et al.[10] whereby a point source located in the transformed medium seems to radiate from a shifted location, to finite size bodies, which undergo the geometric transformation (4.1). As a result, an infinite conducting object placed inside the heterogeneous anisotropic coating of the cloak, appears to an external observer as another infinite conducting one.

**Acknowledgements**: A.D. and S.G. thank the European Research Council for its support through ERC Starting Grant ANAMORPHISM.